\documentclass[conference,10pt]{IEEEtran}
\IEEEoverridecommandlockouts
\usepackage{cite}
\usepackage{amsmath,amssymb,amsfonts}
\usepackage{algorithmic}
\usepackage{graphicx}
\usepackage{textcomp}
\usepackage{inputenc}
\usepackage{braket}
\def\BibTeX{{\rm B\kern-.05em{\sc i\kern-.025em b}\kern-.08em
    T\kern-.1667em\lower.7ex\hbox{E}\kern-.125emX}}
    
\begin{document}

\title{
The Shannon-McMillan Theorem Proves Convergence to Equiprobability
of Boltzmann's Microstates
\\
}

\author{Arnaldo Spalvieri$^*$ \\
$^*$Dipartimento di Elettronica, Informazione e Bioingegneria,
Politecnico di Milano
}

\maketitle

\begin{abstract}
The paper shows that, for large number of particles and for
distinguishable and non-interacting identical particles,
convergence to equiprobability of the $W$ microstates of the
famous Boltzmann-Planck entropy formula $S=k \log(W)$ is proved by
the Shannon-McMillan theorem, a cornerstone of information theory.
This result further strengthens the link between information
theory and statistical mechanics.

\end{abstract}
\section{Introduction}

Entropy plays a key role in communication and information theory
and in thermodynamics. Starting from the concept of entropy, many
authors built in the past bridges between information and
communication theory and thermodynamics, so that all the textbooks
of statistical mechanics and thermal physics have chapters or
sections where the basics of information theory are introduced. A
recent survey about the links between entropy in communications
and in thermodynamics is \cite{mariosurvey}. Making a
comprehensive review of the bibliography on this topic is out of
the scope of the present paper. We limit ourselves to mention the
pioneering work of Jaynes \cite{jaynes2} and the large attention
that has been attracted in the past by the connections between
information theoretic inequalities and the irreversibility of
certain thermodynamical processes, see \cite{merhav}, Chapter 4 of
\cite{cover}, \cite{landauer}.

The aim of this paper is to put the information-theoretic concept
of {\em typical set} inside the framework of statistical
mechanics, thus strengthening the link between the two
disciplines. Typicality is not a new word in statistical
mechanics. It has been introduced in \cite{gold} for system that
are not at the equilibrium to mean that these systems, in their
spontaneous evolution, {\em typically} tend to the maximum
entropy. The difference with \cite{gold} is that (non) equilibrium
is not our concern and that the typicality we refer to is the one
that is technically established in information theory by the
Shannon-McMillan theorem. Specifically, this paper shows that the
Shannon-McMillan theorem {\em proves} that the microstates of the
famous Boltzmann-Plank entropy formula are equiprobable. This
result is surprising because, since the times of Boltzmann, the
entire scientific community has {\em postulated} equiprobability
of microstates. Today, all the most influential textbooks of
statistical mechanics and thermal physics unanimously say that
microstates' equiprobability is the central postulate of
statistical mechanics, see for instance 4.2 of \cite{kardar},
chapter 6 of \cite{huang}, 16.1 of \cite{sek}, 1.1 of
\cite{pathria}. In 3.4 of \cite{fitz}, the author not only
postulates equiprobability, but also expresses, in the passage
that we hereafter quote, strong scepticism about the possibility
of proving it.

{\em It is, unfortunately, impossible to prove with mathematical
rigor that the principle of equal a priori probabilities applies
to many-particle systems. Over the years, many people have
attempted this proof, and all have failed. Not surprisingly,
therefore, statistical mechanics was greeted with a great deal of
scepticism when it was first proposed in the late 1800s. One of
the its main proponents, Ludwig Boltzmann, became so discouraged
by all of the criticism that he eventually committed suicide.
Nowadays, statistical mechanics is completely accepted into the
cannon of physics - quite simply because it works.}

The outline of the paper is as follows. In Section II we introduce
our model of the thermodynamical system, where the only assumption
that we make is that the system is made by a large number of
non-interacting and identical particles. We say {\em our model}
because it is slightly different from the commonly accepted one
even if, in the thermodynamical limit of infinite number of
particles, our model becomes compatible with the standard one.
Section III shows that, with the further assumption of
distinguishable particles, the set of microstates accessible to
the thermodynamical system is the information-theoretic typical
set of the Shannon-McMillan theorem. A crucial property of the
typical set is that the probability distribution of its element
converges to uniformity, therefore the identification between the
typical set and the set of accessible microstates proves
convergence to equiprobability of the accessible microstates.
Finally, in Section IV we recap the main points of the paper and
draw the conclusions.

\section{Notation and System Model}
Let the uppercase calligraphic character, e.g. ${\cal X}$, denote
discrete random variables, let the set $\{ {\cal X} \}$ be the
support set of the random variable, e.g.
\[\{{\cal X}\}=\{x_0,x_1,\cdots\}\]
and let $|{\cal X}|$ be the number of elements of the support set.
The probability $Pr({\cal X}=x)$ of the event ${\cal X}=x$ is
denoted with the shorthand $p(x)$,
\[Pr({\cal X}=x)=p(x), \ \ \forall \ x \in \{{\cal X}\}.\]
The probability distribution of ${\cal X}$ is the deterministic
function $p(\cdot)$ of the random ${\cal X}$ with
\[\{p({\cal X})\}=\{p(x_0),p(x_1),\cdots\}\]
and
\[Pr(p({\cal X})=p(x))=p(x), \ \
\forall \ \ x \in \{{\cal X}\}.\] Consider a thermodynamical
system made by a fixed number $N$ of non-interacting identical
particles with one degree of freedom each. The thermodynamical
system is modelled here by the random vector
\[{\cal E}_1^N=({\cal E}_1, {\cal E}_2, \cdots, {\cal E}_N)\]
of the energies of the $N$ particles. We assume that energy is
quantized and that the random entries of the random vector are
independent (non interacting) identically distributed (identical)
(i.i.d.) random variables (particles), that is
\begin{equation}p({\cal E}_1^N)=\prod_{i=1}^{N}p({\cal
E}_i),\label{iid}\end{equation} with
\[p({\cal E}_1)=p({\cal E}_2)=\cdots=p({\cal E}_N).\]
Since particles are i.i.d., one of them represents all. The random
energy of the generic particle, that we call ${\cal E}$, takes its
values in the set $\{{\cal E}\}$ of the allowed energy levels
\begin{equation}\{{\cal E}\}=\{\epsilon_0,\epsilon_1,\cdots \}.\label{support}
\end{equation} The expected energy of the particle is
\begin{equation}
E=\sum_{\epsilon \in \{{\cal E}\}} \epsilon p(\epsilon). \nonumber
\end{equation}
The total random energy of the system is
\begin{equation}\sum_{i=1}^N {\cal E}_i.
\nonumber
\end{equation} By the i.i.d. assumption and by the
law of large numbers, when $N \rightarrow \infty$, the system's
random energy divided by the number of particles, that is, the
random mean energy, tends to the deterministic expected energy:
\begin{equation}\lim_{N \rightarrow \infty}\frac{1}{N} \sum_{i=1}^N {\cal E}_i
= E .\label{borele}\end{equation}  Convergence of the limit can be
in various senses. In this paper, we will focus our attention on
the weak law of large numbers (Khinchin law), hence we consider
convergence of the limit (\ref{borele}) {\em in probability},
meaning that, for every $\eta
>0$,
\begin{equation} \lim_{N \rightarrow \infty}Pr
\left(\left|\frac{1}{N}\sum_{i=1}^N {\cal E}_i- E \right|> \eta
\right) = 0.\label{conv}
\end{equation}
The limit in (\ref{conv}) is interpreted by saying that, for every
$\eta$, if $N$ is sufficiently large, then the total energy of the
system lies in a narrow interval $\pm N \eta$ around $N E$ with
high probability.

A comment is in order about the approach pursued in this paper,
that is based on the expected energy per particle, in comparison
to the classical approach of thermal physics and statistical
mechanics where the total system energy is considered. The
assumption that system's total energy lies in a narrow interval is
the basis of the classical {\em microcanonical ensemble} approach,
see for instance chapters 1 and 2 of \cite{pathria}. In the
microcanonical ensemble approach, the system is assumed to have a
large and fixed number of particles and to be closed. Since $N$ is
large and fixed, basically there is no difference between the
expected energy per particle and the total energy divided by the
number of particles. When the number of particles is fixed but not
so large, the classical approach in statistical mechanics and
thermal physics is that of the {\em canonical ensemble}, see for
instance chapter 3 of \cite{pathria}. In the canonical ensemble
approach, the thermodynamical system is at the thermal equilibrium
at a temperature $T$ with a heat bath and interacts with it to
maintain the equilibrium condition. One of the consequences of the
interaction between system and heat bath is that the total energy
of the system fluctuates, hence it becomes {\em substantially}
random. Due to the random nature of total system's energy, in the
canonical ensemble approach it can be no more used to characterize
the system, which is characterized instead by the temperature of
the heat bath. While system's total energy becomes substantially
random and must be abandoned in favor of the temperature of the
heat bath, the expected energy, which is deterministic, can be
related in a deterministic way to the temperature of the heat bath
by the appropriate energy-temperature relation. For instance, the
high temperature approximation for wholly kinetic energy is
\begin{equation}E= \frac{kT}{2},
\nonumber
\end{equation}
where $T$ is in Kelvin degrees and $k$ is Boltzmann's constant.
Hence, although in the following we will limit ourselves to $ N
\rightarrow \infty$, we see that the expected energy per particle
can be used to characterize both the microcanonical and the
canonical ensembles.

\section{Equiprobability of Accessible Microstates}

When we become aware of the result of a random experiment, in our
case, the energy level of a particle, the {\em surprise} that we
experience is the following deterministic function of the random
result:
\begin{align}-\log(p({\cal
E})), \nonumber 
\end{align} where $\log(x)$
indicates the logarithm of $x$ and the base of the logarithm
depends on the context. In information theory the base is 2 while
in physics the base is Euler's constant $e$.  Being a function of
random variable, the surprise is itself a random variable. Its
expectation is Shannon's entropy, that we call $H$:
\begin{equation}H=-\sum_{\epsilon \in \{{\cal
E}\}}p(\epsilon)\log(p(\epsilon)).\label{shannon}\end{equation}

 If distinguishable
particles, or indexed particles, are considered, an outcome of
${\cal E}_1^N$ is an {\em energy microstate}, or, simply, {\em
microstate}. For instance, a microstate is
\begin{equation}
{\cal E}_1^N: \ {\cal E}_1=\epsilon_5, \  {\cal
E}_2=\epsilon_{31}, \ \cdots,  \ {\cal
E}_N=\epsilon_{18},\label{example}
\end{equation} meaning that the energy of the first particle is
$\epsilon_5$, the energy of the second particle is
$\epsilon_{31}$, $\cdots$, the energy of the $N$-th particle is
$\epsilon_{18}$. This definition of microstate matches the
commonly accepted one, see e.g. 1.1 of \cite{pathria}, where the
microstates of the system are the independent solutions of the
Schr\"oedinger equation of the system whose eigenvalue is the
total energy of the system. Due to the assumption of independency
between particles, the system decouples into $N$ independent
equations whose individual solutions are the wave functions whose
energy eigenvalues are the energy levels (\ref{support}), the sum
of the individual energy eigenvalues being equal to system's total
energy. By now we settle for the definition of microstate, we will
return soon on the concept of accessible microstate and on the
energy constraint that the system must comply with.

When we become aware of the microstate visited by the system, the
surprise that we experience is
\begin{align}-\log(p({\cal
E}_1^N)). \nonumber
\end{align} The expectation of
the above surprise is the Shannon entropy of the system, hence the
Shannon entropy of the random vector ${\cal E}_1^N$:
\begin{equation}-\sum_{\epsilon_1^N \in \{{\cal
E}_1^N\}}p(\epsilon_1^N)\log(p(\epsilon_1^N)).\label{gibbs}\end{equation}
In statistical mechanics, (\ref{gibbs}) is called {\em Gibbs
entropy}. By the i.i.d. assumption (\ref{iid}), the entropy
(\ref{gibbs}) is equal to $n$ times the Shannon entropy of one
particle:
\begin{equation}-\sum_{\epsilon_1^N \in \{{\cal
E}_1^N\}}p(\epsilon_1^N)\log(p(\epsilon_1^N))=NH,\nonumber\end{equation}
see 2.6.6 of \cite{cover}. One immediately recognizes that
\begin{align} \lim_{N \rightarrow \infty }
-\frac{1}{N}\log(p({\cal E}_1^N)) &= \lim_{N \rightarrow \infty }
- \frac{1}{N}\sum_{i=1}^N\log(p({\cal E}_i))\label{inda}
 \\ & = H,\label{syse}\end{align} where (\ref{inda}) is the
assumption of independency between particles and (\ref{syse})
follows from the assumption of identically distributed random
variables and from the law of large numbers. The equality between
the leftmost term and the rightmost one in the above two
equalities shows that, as $N \rightarrow \infty$, the distribution
of the microstates that the system can visit converges to the
uniform distribution, because $H$ is a deterministic and fixed
quantity, independent of the specific microstate ${\cal E}_1^N$.
The convergence of the limit (\ref{inda}) to (\ref{syse}) is in
various senses. Convergence in probability, that is
\begin{equation} \lim_{N \rightarrow \infty}Pr\left(\left|\frac{1}{N}
\log( p({\cal E}_1^N)))
 + H \right|> \eta \right) = 0,\label{conv2} \end{equation}
is referred to as the Asymptotic Equipartition Property (AEP) in
chapter 3 of \cite{cover},\footnote{Here it is the probability
that is equally partitioned, not the energy, hence this AEP has
nothing to do with the classical energy equipartition property.}
while many authors refer to (\ref{conv2}) as to the
Shannon-McMillan theorem. Almost everywhere convergence, that
strengthens convergence in probability, is known as the
Shannon-McMillan-Breiman theorem \cite{bre}. Exactly as it happens
with total system's energy, here, for sufficiently large but
finite $N$, the surprise about system's microstate lies in a
narrow range $\pm N \eta$ around $NH$ with high probability and,
exactly as it happens with the expected energy per particle $E$,
$H$, which is {\em per particle}, can be used to characterize the
entropy both in the microcanonical and the in canonical ensemble
approaches even if, in what follows, we will limit ourselves to $N
\rightarrow \infty$.

\subsection{Typical Set and Accessible Microstates}


Starting from (\ref{conv2}) we arrive to the definition  of {\em
typical set}, which is the subset $\{{\cal T}_1^N(\eta)\}$ of $
\{{\cal E}_1^N\}$ made by the vectors ${\cal E}_1^N$ whose
probability, for any $\eta$ and for sufficiently large $N$, is in
the narrow range
\begin{equation}e^{-N(H+\eta)} \leq p({\cal
E}_1^N) \leq e^{-N(H-\eta)}.\label{narrow}
\end{equation}
The properties of the typical set are that the number of its
elements is in the narrow range
\begin{equation}e^{N(H+\eta)} \geq |{\cal T}_1^N(\eta)|
\geq (1-\eta)e^{N(H-\eta)}\label{narrow2}
\end{equation}
and that the probability that the system visits one element of the
typical set is
\begin{equation}Pr({\cal E}_1^N \in \{{\cal T}_1^N(\eta)\})>
1-\eta, \label{narrow3}
\end{equation}
see again \cite{cover}.
 Formulas (\ref{narrow})-(\ref{narrow3}) show
that with high probability the system visits, or, in the language
of statistical mechanics, {\em makes access to}, one among $|{\cal
T}_1^N(\eta)|$ microstates. Also, the distribution of microstates
converges to the uniform one as $N \rightarrow \infty$,  because
$|{\cal T}_1^N(\eta)|$ is basically the inverse of $p({\cal
E}_1^N)$, which is constrained to lie in the narrow range
(\ref{narrow}).
 Note that convergence to equiprobability of accessible microstates
is a {\em consequence} of the i.i.d. assumption and of $N
\rightarrow \infty$, hence, in the end, of the law of large
numbers, not a {\em postulate}.

The AEP divides the support set of energy microstates into two
subsets: the typical set, that is the set of accessible
microstates, and its complement, that is the set of microstates
that the system cannot access. The division can be appreciated by
observing that
\[|{\cal E}| - e^{H} \geq 0,\]
 with equality when the energy levels that the single
particle can visit are equiprobable, hence \begin{equation}|{\cal
E}_1^N|=|{\cal E}|^N \geq e^{NH}.\label{ineq}
\end{equation} Writing, with some abuse of mathematics,
(\ref{narrow2}) as
\[|{\cal T}_1^N(\eta)| \approx e^{NH},\] we see that,
when inequality (\ref{ineq}) fits with equality, the size of the
typical set is the entire support set made of $|{\cal E}|^N$
elements, while, when $|{\cal E}|
> e^{H}$, the number of elements of the typical set is
lower than $|{\cal E}|^N$, leading to the division of the support
set into the two subsets mentioned above. In this case, the system
will visit with high probability the typical set and will visit
with vanishingly small probability the complement of the typical
set to the support set.

We now return on accessibility and on the energy constraint to
comment again on the key difference between our approach and the
standard one. In the standard approach, for a microstate to be
accessible, the energy eigenvalue of the wave function that solves
system's Schr\"oedinger equation of the system must be equal to
the constraint of total energy that is imposed on the system,
while here it is the probability distribution $\{p({\cal E})\}$
that must have a prescribed expected energy per particle.
Compatibility between our approach and the standard one is
guaranteed because, when $N \rightarrow \infty$, the energy
eigenvalue of the system divided by the number of particles {\em
is forced} by the law of large numbers to be equal to the expected
energy of one particle, see (\ref{borele}).

\subsection{Boltzmann-Plank entropy}

The equality between the Boltzmann-Plank entropy and $N$ times the
Shannon entropy of one particle is widely accepted in textbooks of
statistical mechanics and thermal physics, see e.g. eqn. 18.14 of
\cite{sek}. However, while in the standard approach both the
Boltzmann-Plank entropy and the Shannon entropy are purely
deterministic quantities, in our approach the Boltzmann-Planck
entropy is the random quantity that we are going to introduce in
(\ref{rbpe}), that, divided by the number of particles, becomes
equal to the deterministic Shannon entropy when the number of
particles tends to infinity. We hereafter discuss the key passage,
the passage where our model becomes compatible with the standard
combinatorial approach to the evaluation of the number of
accessible microstates.

Let ${\cal N}(\epsilon)$ be the random number of particles that
visit the generic energy level $\epsilon$, with
\begin{equation}\sum_{\epsilon \in \{{\cal E}\}}{\cal N}(\epsilon)=N.\label{totaln}\end{equation}
Compatibility of the accessible microstates with constraints
imposed on the system will be guaranteed by imposing
compatibility of the set
\begin{equation}\{{\cal N}(\epsilon_0), {\cal N}(\epsilon_1),
\cdots\}\label{neps}\end{equation}  with the distribution
$\{p({\cal E})\}$. The random number of microstates is the
multinomial coefficient
\begin{equation}
{\cal W}=\frac{N!}{\prod_{\epsilon \in \{{\cal E}\}}{\cal
N}(\epsilon)!}\label{w} \end{equation} and the random
Boltzmann-Plank entropy is
\begin{equation}
{\cal S}=k \log({\cal W})\label{rbpe}
\end{equation}
Randomness of the set (\ref{neps}) and, as a consequence, of the
number of microstates (\ref{w}) and of the Boltzmann-Plank entropy
(\ref{rbpe}), marks the difference between our approach and the
standard, deterministic, one. Exactly as in the case of the
energy, once again the law of large numbers bridges the gap
between randomness and determinism, thus making our {\em random}
approach compatible with the {\em deterministic} standard
approach, hence the key passage is\footnote{Specifically, step by
step one has
\begin{align}& \lim_{N \rightarrow \infty} \frac{1}{k N}
{\cal S}=  \lim_{N \rightarrow \infty} \frac{1}{N} \log ({\cal W})
\nonumber \\ & = \lim_{N \rightarrow \infty}
\frac{1}{N}\left(\log(N!)-\sum_{{\epsilon \in \{{\cal
E}\}}}\log({\cal N}(\epsilon)!) \right) \nonumber
\\  &= \lim_{N \rightarrow
\infty}\frac{1}{N}\left(N\log\left(\frac{N}{e}\right)
-\sum_{{{\epsilon \in \{{\cal E}\}}}}{\cal
N}(\epsilon)\log\left(\frac{{\cal N}(\epsilon)}{e}\right)\right)
\label{stirling}
\\ &= \lim_{N \rightarrow \infty}\frac{1}{N}\left(N\log(N) -
\sum_{{{\epsilon \in \{{\cal E}\}}}}{\cal N}(\epsilon)\log({\cal
N}(\epsilon)) \right)\label{sat1}
\\ &= \lim_{N \rightarrow
\infty}\sum_{{{\epsilon \in \{{\cal E}\}}}}\frac{{\cal
N}(\epsilon)}{N}\left(\log(N) -
\log({\cal N}(\epsilon)) \right)\label{sat2}\\
&=
 \lim_{N \rightarrow \infty}
-\sum_{{{\epsilon \in \{{\cal E}\}}}}\frac{{\cal
N}(\epsilon)}{N}\log\left(\frac{{\cal N}(\epsilon)}{N}\right)
\nonumber
\\ & = -\sum_{\epsilon \in \{{\cal
E}\}}p(\epsilon)\log(p(\epsilon)) \label{lln} \\ &= H \nonumber,
\end{align}  where in (\ref{sat1}) and (\ref{sat2}) we use
(\ref{totaln}), (\ref{lln}) is the law of large numbers and
(\ref{stirling}) is Stirling's formula that, in the big ${\cal O}$
notation, is
\[\log (N!) =N \log\left(\frac{N}{e}\right) + {\cal O}(\log(N)),\]
and we put in (\ref{stirling})
\[\lim_{N \rightarrow \infty} \frac{{\cal O}(\log(N))}{N}=
\lim_{N \rightarrow \infty} \frac{{\cal O}(\log({\cal
N}(\epsilon)))}{N} =0, \ \ \forall \ \epsilon \in \{{\cal E}\} .\]
The law of large number is also invoked in \cite{frigg} in a
passage that is basically the same as our (\ref{lln}) to show the
equivalence of the phase space formulation of Gibbs and
Boltzmann-Plank entropies.}
\begin{equation}
\lim_{N \rightarrow \infty}\frac{1}{N}\log({\cal W})=H.\label{key}
\end{equation}
From (\ref{narrow2}) and (\ref{key}) we conclude that, for every
small $\eta >0$,
\begin{equation}
\lim_{N \rightarrow \infty}\frac{1}{N}\log({\cal W})=\lim_{N
\rightarrow \infty}\frac{1}{N}\log(|{\cal
T}_1^N(\eta)|).\label{conclusion}
\end{equation}   Therefore, not only our definition of energy
microstates exemplified in (\ref{example}) is the commonly
accepted one, but also the logarithm of the number of accessible
microstates that we obtain by our approach is, for $N \rightarrow
\infty$, equal to the logarithm of the standard combinatorial
definition (\ref{w}) of accessible microstates.

\section{Conclusion}
The paper has demonstrated that there is no need of {\em assuming}
or {\em postulating} equiprobability of accessible microstates,
because convergence of microstates' distribution to uniformity is
a consequence of the large number of particles, of the i.i.d.
assumption and of the assumption of distinguishable particles.
Specifically, equations (\ref{narrow})-(\ref{narrow3}) show that
convergence to equiprobability is a consequence of the AEP and
equation (\ref{conclusion}) shows that the microstates we are
dealing with are the microstates that are counted by the standard
combinatorial approach.

It is worth observing that equiprobability has nothing to do with
the thermal equilibrium, hence with entropy maximization.
Actually, the only conditions for convergence to equiprobability
are the assumptions that we made, which do not include thermal
equilibrium and/or entropy maximization, therefore the
distribution converges to the uniform one also if the system is
not at the thermal equilibrium. What can change between a system
at the equilibrium and a system that is not at the equilibrium
when both are subject to the same constraints is the probability
distribution of energy levels and, with it, Shannon's entropy and,
with it again, the number of accessible microstates, which,
however, tend to equiprobability. Since the accessible microstates
remain virtually equiprobable also in systems that are not at the
equilibrium, the entropy of these systems is still given by the
Boltzmann-Plank entropy formula, provided that the number of
accessible microstates is conveniently expressed through the
Shannon entropy of the specific (non-equilibrium) probability
distribution of the energy of the individual particle by
(\ref{shannon}), (\ref{narrow2}).

 When $N$ is not large, e.g. $N=1$, microstates
are in general not equiprobable, independently of the entropy of
the system.

To summarize, the main result of this paper is:
\begin{itemize}
\item for large $N$ and i.i.d. particles, the accessible
microstates are always virtually equiprobable, also for systems
that are not at the thermal equilibrium; \item for small $N$, the
accessible microstates are never equiprobable, also for systems
that are at the thermal equilibrium with a heat bath.
\end{itemize}

The important case of indistinguishable particles is not treated
in the paper and is left to future research.



\begin{thebibliography}{999}

\bibitem{mariosurvey}  Martinelli, M. "Photons, Bits and Entropy: From Planck to Shannon
at the Roots of the Information Age." {\em Entropy}, 2017, {\em
19}, 341.

\bibitem{jaynes2} Jaynes, E. T. "Information theory and
statistical mechanics." {\em Physical review}, 1957, {\em 4},
620--630.


\bibitem{merhav}  Merhav, N. "Physics of the Shannon Limits,"
{\em IEEE Trans. on Inform. Theory,} 2010, {\em 9}, 4274--4285.

\bibitem{cover} Cover, T. M.; and Thomas, J. A.
"Elements of information theory, 2$^{nd} $ edition."  Wiley series
in telecommunications and signal processing, 2006.


\bibitem{landauer} Landauer, R. "Irreversibility and heat generation in the
computing process." IBM journal of research and development 5.3
(1961): 183-191.

\bibitem{gold}
Goldstein, Sheldon, and Joel L. Lebowitz. "On the (Boltzmann)
entropy of non-equilibrium systems," 2004, Physica D: Nonlinear
Phenomena 193.1-4, 53--66.



\bibitem{kardar}
Kardar, M. "Statistical Physics of Particles." Cambridge
University Press, 2007.

\bibitem{huang}
Huang, K. "Statistical Mechanics." Wiley, 1987.

\bibitem{sek}
Sekerka, R. F. "Thermal Physics: Thermodynamics and Statistical
Mechanics for Scientists and Engineers." Elsevier, 2015.

\bibitem{pathria} Pathria, R. K.; Beale, P. D. "Statistical Mechanics,
3$^{rd} $ edition." Elsevier, 2011.



\bibitem{fitz} Fitzpatrick, R. "Thermodynamics and Statistical Mechanics."
World Scientific, 2020.

\bibitem{bre} Breiman, L. "The individual ergodic theorem of information theory."
The Annals of Mathematical Statistics 28.3 (1957): 809-811.

\bibitem{frigg}
Frigg, R.; Werndl, C. "A guide for the perplexed." Probabilities
in physics 2 (2011): 115.


%


%
%
%
%
%
%
%
%
%
%
%
%
%
%
%
%
%
%
%
%
%
%
%
%
%
%
%
%
%
%

%
%
%
%
%
%
%
%


%
%
%
%
%
%
%
%
\end{thebibliography}
\end{document}